\documentclass[a4paper,10pt,eqno]{article}
\usepackage{theorem}
\usepackage{latexsym,amssymb,amsfonts,amsmath}
\newtheorem{thm}{Theorem}[section]
\newtheorem{lem}[thm]{Lemma}
\newtheorem{cor}[thm]{Corollary}

\theoremheaderfont{\scshape}

\newcommand{\ZM}{\mathbb{Z}}

\newcommand{\CM}{\mathbb{C}}

\newcommand{\ket}[1]{|#1\rangle}

\title{{\Large {\bf Return probability of one-dimensional discrete-time \\ 
quantum walks with final-time dependence}}}
\author{
{\small Yusuke Ide\footnote{To whom correspondence should be addressed. E-mail: ide@kanagawa-u.ac.jp}}\\
{\scriptsize Department of Information Systems Creation, 
Faculty of Engineering, 
Kanagawa University}\\
{\scriptsize Kanagawa, Yokohama 221-8686, Japan}\\
{\scriptsize e-mail: ide@kanagawa-u.ac.jp}\\
{\small Norio Konno}\\
{\scriptsize Department of Applied Mathematics, 
Faculty of Engineering, 
Yokohama National University}\\
{\scriptsize Hodogaya, Yokohama 240-8501, Japan}\\
{\scriptsize e-mail: konno@ynu.ac.jp}\\
{\small Takuya Machida}\\
{\scriptsize Meiji Institute for Advanced Study of Mathematical Sciences,  
Meiji University}\\
{\scriptsize Tama, Kawasaki 214-8571, Japan}\\
{\scriptsize e-mail: bunchin@meiji.ac.jp}\\
{\small Etsuo Segawa}\\
{\scriptsize Department of Mathematical Informatics, 
The University of Tokyo}\\
{\scriptsize Bunkyo, Tokyo 113-8656, Japan}\\
{\scriptsize e-mail: segawa@stat.t.u-tokyo.ac.jp}\\
}
\date{\empty }
\begin{document}
\maketitle

\par\noindent
\begin{small}
\par\noindent
{\bf Abstract}. We analyze final-time dependent discrete-time quantum walks in one dimension. We compute asymptotics of the return probability of the quantum walk by a path counting approach. Moreover, we discuss a relation between the quantum walk and the corresponding final-time dependent classical random walk.

\footnote[0]{
{\it Abbr. title:} Return probability of quantum walks with final-time dependence
}
\end{small}

\setcounter{equation}{0}
\section{Introduction}
The discrete-time quantum walk (QW) was first intensively studied by Ambainis et al.\ \cite{AmbainisEtAl2001} and has been widely studied for a decade as a quantum counterpart of the random walk. There are reviews, for example Kempe \cite{Kempe2003}, Kendon \cite{Kendon2007}, Venegas-Andraca \cite{VAndraca2008}, Konno \cite{Konno2008b}. In this paper we consider the final-time dependent discrete-time QW determined by a sequence of $2 \times 2$ unitary matrices. Particularly, we focus on the return probability which is the probability that the walker returns to the origin. As for the work on the return probability on QWs, see \v{S}tefa\v{n}\'ak et al.\ \cite{StefanakEtAl2008a,StefanakEtAl2008b,StefanakEtAl2009}, for example.

Mackay et al. \cite{MackayEtAl2002} and Ribeiro et al. \cite{RibeiroEtAl2004} studied random unitary sequences and suggested numerically that the probability distribution of the QW converges to a binomial distribution by averaging over many trials. Their result was proved by Konno \cite{Konno2005b}. Ribeiro et al. \cite{RibeiroEtAl2004} also reported that periodic sequence is ballistic and Fibonacci one is sub-ballistic by numerical evaluations. Machida and Konno \cite{MachidaKonno2009} presented weak limit theorems for time-dependent QWs by the Fourier analysis. Romanelli \cite{Romanelli2009} introduced a time-dependent QW and found an analytical expression for the asymptotics of the standard deviation by a continuum limit. This expression yields that the parameter $\alpha$ (defined by Eq. (\ref{Roma}) in this paper) determines the degree of the spread of the QW for large time. The result is consistent with the numerical result. 

Related to the return probability, localization is one of important topics in the study of QW. We say that a model exhibits localization when a long-time limit of the return probability remains positive. For the time-independent and space-homogeneous case, Inui et al.\ \cite{InuiEtAl2005} first showed that a three-state QW on $\ZM=\{0, \pm 1, \pm 2, \ldots \}$ exhibits localization. For the time-independent but space-inhomogeneous case, indeed the Hadamard walk with one defect, Konno \cite{Konno2010} proved that localization occurs for a two-state QW on $\ZM$. Furthermore, Cantero et al.\ \cite{CanteroEtAl2010} provided explicit expressions for the asymptotic return probabilities for a more general setting. 
In the present paper, we analyze the return probability of the QW, which has a connection with Romanelli \cite{Romanelli2009}, and the related QW. Furthermore, we compute the probability of the corresponding final-time dependent classical random walk (CRW). 

The rest of the paper is organized as follows. Section 2 treats the definition of the model. In Sect. 3, we present main results (Theorems {\rmfamily \ref{thm1}} and {\rmfamily \ref{thm2}}) of this paper. Section 4 is devoted to proofs of both theorems. The results of the corresponding classical model are given in Sect. 5. We give the proofs of these results in Sect. 6 and finally in Sect. 7, we discuss our results.

\section{Definition of the walk}
In this section, we give the definition of a two-state QW with final-time dependence on $\ZM$ considered here, where $\ZM$ is the set of integers. Let $\ZM_{+} = \{1,2, \ldots \}$ be the set of positive integers. For the general setting, the time evolution of the walk is determined by a sequence of $2 \times 2$ unitary matrices, $\{ U_n : n \in \ZM_{+} \}$, where 
\begin{align*}
U_n =
\left[
\begin{array}{cc}
a_n & b_n \\
c_n & d_n
\end{array}
\right],
\end{align*}
with $a_n, b_n, c_n, d_n \in \mathbb C$ and $\CM$ is the set of complex numbers. The subscript $n$ means the final time. 

The discrete-time QW is a quantum version of the classical random walk with additional degree of freedom called chirality. The chirality takes values left and right, and it means the direction of the motion of the walker. At each time step, if the walker has the left chirality, it moves one step to the left, and if it has the right chirality, it moves one step to the right. Let define
\begin{eqnarray*}
\ket{L} = 
\left[
\begin{array}{cc}
1 \\
0  
\end{array}
\right],
\qquad
\ket{R} = 
\left[
\begin{array}{cc}
0 \\
1  
\end{array}
\right],
\end{eqnarray*}
where $L$ and $R$ refer to the left and right chirality state, respectively.  To define the dynamics of our model, we divide $U_n$ into two matrices:
\begin{eqnarray*}
P_n =
\left[
\begin{array}{cc}
a_n & b_n \\
0 & 0 
\end{array}
\right], 
\quad
Q_n=
\left[
\begin{array}{cc}
0 & 0 \\
c_n & d_n 
\end{array}
\right],
\end{eqnarray*}
with $U_n=P_n+Q_n$. The important point is that $P_n$ (resp. $Q_n$) represents that the walker moves to the left (resp. right) at position $x$ at each time step.

For fixed the final time $n$, we let $\Xi^{(n)} _{k} (l,m)$ denote the sum of all paths starting from the origin in the trajectory consisting of $l$ steps left and $m$ steps right at time $k (\le n)$ with $l+m=k$. For example, when $n \ge 3$, we have
\begin{align*}
\Xi^{(n)} _2 (1,1) = Q_n P_n + P_n Q_n, \quad \Xi^{(n)} _3 (2,1) = Q_n P_n^2 + P_n Q_n P_n + P_n^2 Q_n.
\end{align*}
Remark that $\Xi^{(n)} _{k} (l,m)$ $(1\leq k \leq n)$ is determined by not $U_{1}, U_{2}, \ldots ,U_{k}$ but only $U_{n}$. Let $X_{k}$ be the position of our quantum walker at time $k$. 
The probability that the walker is in position $x$ at time $k (\le n)$ starting from the origin with the initial qubit state $\varphi_{\ast}$ is defined by 
\begin{align*}
P (X_{k} =x) = || \Xi^{(n)}_{k}(l, m) \varphi_{\ast} ||^2,
\end{align*}
where $k=l+m$, $x=-l+m$ and $\varphi_{\ast}$ is a linear combination of $\ket{L}$ and $\ket{R}$ with $||\varphi_{\ast}||^{2}=1$. In particular, we take $\varphi_{\ast} = {}^T [1/\sqrt{2},i/\sqrt{2}]$, where $T$ is the transposed operator. Then the probability distribution of the walk is symmetric. The following is the important quantity of this paper.
\begin{align*} 
p_n (0) = P (X_{n} =0).
\end{align*}
This is the return probability at the time $n$. Therefore, $p_{2n+1} (0) = 0$ for $n \ge 0$. 

In the present paper, we concentrate on the unitary matrix $U^{R}_n$ with fixed parameters $\alpha, \tau > 0$, which was recently introduced and studied by Romanelli \cite{Romanelli2009}: 
\begin{align}
U^{R}_n =
\left[
\begin{array}{cc}
\left( \frac{\tau}{n} \right)^{\alpha}  &  \sqrt{1 - \left( \frac{\tau}{n} \right)^{2 \alpha} } \\
\sqrt{1 - \left( \frac{\tau}{n} \right)^{2 \alpha} } & - \left( \frac{\tau}{n} \right)^{\alpha}
\end{array}
\right].
\label{Roma}
\end{align}
However, the evolution of his model is different from that of our QW. Because for fixed final time $n$, his QW at time $k \in \{1,2,\ldots, n\}$ is determined by $U_{k}^{R}$. On the other hand, that of our QW at time $k \in \{1,2,\ldots, n\}$ is given by $U_{n}^{R}$ which depends only on the final time $n$. 
Moreover we consider the following $U^{K}_n$ with $\alpha, \tau > 0$: 
\begin{align*}
U^{K}_n =
\left[
\begin{array}{cc}
\sqrt{1 - \left( \frac{\tau}{n} \right)^{2 \alpha} } &  \left( \frac{\tau}{n} \right)^{\alpha} \\
\left( \frac{\tau}{n} \right)^{\alpha} & - \sqrt{1 - \left( \frac{\tau}{n} \right)^{2 \alpha} }
\end{array}
\right].
\end{align*}
The two models defined by the unitary matrices $U_{n}^{R}$ and $U_{n}^{K}$ seem to be similar, but as we will show in the next section, asymptotic behaviors of the return probability are different. Furthermore, if $U_n ^{R}$ with $\alpha =0$, then $p_{2n} (0)=0$ for any $n \ge 1.$ On the other hand, if $U_n ^{K}$ with $\alpha =0$, then $p_{2n} (0)=1$ for any $n \ge 0.$ For a time-independent QW, the Hadamard walk has been extensively investigated by many researchers, which is defined by $U_1 ^{R}$ (or equivalently, $U_1 ^{K}$) with $\alpha =1$ and $\tau = 1/\sqrt{2}.$ In the symmetric Hadamard walk, it is known that 
\begin{align*}
\lim_{n \to \infty} n \> p_{2n} (0) = \frac{1}{\pi}, 
\end{align*}
see Konno \cite{Konno2010b,Konno2010}, for example. For the corresponding symmetric classical random walk, 
\begin{align*}
\lim_{n \to \infty} \sqrt{n} \> p_{2n} (0) = \frac{1}{\sqrt{\pi}}. 
\end{align*}


\section{Our results}
In this section, we give results on asymptotics of the return probability of final-time dependent 1D QWs determined by $U_n ^{R}$ and $U_n ^{K}$, respectively.

\begin{thm}
\label{thm1}
For final-time dependent 1D QW determined by $U_n ^{R}$ with $\alpha, \tau >0$, we have 
\begin{align*}
p_{2n} (0) \sim J_0 (\tau^{\alpha} (2n)^{1 - \alpha})^2  + J_1 (\tau^{\alpha} (2n)^{1 - \alpha})^2,
\end{align*}
where $p_{2n} (0)$ is the return probability at the origin at the final time $2n$, $J_k(x)$ is the Bessel function of the first kind of order $k$, and $f(n) \sim g(n)$ means $f(n)/g(n) \to 1 \> (n \to \infty)$.
\end{thm}
The proof of this theorem will appear in the next section. By using the result, we have 
\begin{cor}
\label{cor1}
(i) If $0 < \alpha < 1$, then 
\begin{align*}
\lim_{n \to \infty} n^{1- \alpha} \> p_{2n} (0) = \frac{1}{\pi} \> \left( \frac{2}{\tau} \right)^{\alpha}. 
\end{align*}
(ii) If $\alpha =1$, then 
\begin{align*}
\lim_{n \to \infty} p_{2n} (0) = J_0 (\tau)^2  + J_1 (\tau)^2.
\end{align*}
(iii) If $\alpha >1$, then 
\begin{align*}
\lim_{n \to \infty} n^{2 (\alpha -1)} \> \{ 1 - p_{2n} (0) \} = \left( \frac{\tau}{2} \right)^{2 \alpha}. 
\end{align*}
\end{cor}
In fact, this corollary $(\alpha \neq 1)$ corresponds to Eq.\ (21)\ in \cite{Romanelli2009} despite the difference of two models. However, it is not clear whether the corollary $(\alpha =1)$ corresponds to his result. Because our result does not have $\ln n$ term. 

From now on we will give a proof of this corollary. As for part (i), we note that as $x \to \infty$, 
\begin{align*}
J_0 (x) \sim \sqrt{\frac{2}{\pi x}} \> \cos (x - \pi/4), \quad J_1 (x) \sim \sqrt{\frac{2}{\pi x}} \> \sin (x - \pi/4),
\end{align*}
see Watson \cite{Watson1944}. Thus $J_0 (x)^2 + J_1 (x)^2 \sim 2/\pi x$. Then we have the desired conclusion. Part (ii) is a direct consequence of the theorem. Concerning part (iii), we see that for small $x$, 
\begin{align*}
J_0 (x) \sim 1 - \frac{x^2}{4}, \quad J_1 (x) \sim \frac{x}{2} - \frac{x^3}{16},
\end{align*}
see \cite{Watson1944}. So we have $J_0 (x)^2 + J_1 (x)^2 \sim 1 - x^2/4$. From this, the proof of part (iii) is complete. 

Next we consider the QW given by $U_{n}^{K}$.

\begin{thm}
\label{thm2}
For time-dependent 1D QW determined by $U_n ^{K}$ with $\alpha, \tau >0$, we have 
\begin{align*}
p_{2n} (0) \sim \left( \frac{\tau}{2n} \right)^{2 \alpha} \> \left\{ J_0 (\tau^{\alpha} (2n)^{1 - \alpha})^2  + J_1 (\tau^{\alpha} (2n)^{1 - \alpha})^2 \right\}.
\end{align*}
\end{thm}
The proof of this theorem can be found in the next section. From this theorem, we obtain the following result as in Corollary {\rmfamily \ref{cor1}}. So we will omit the proof.
\begin{cor}
\label{cor2}
(i) If $0 < \alpha < 1$, then
\begin{align*}
\lim_{n \to \infty} n^{\alpha +1} \> p_{2n} (0) = \frac{1}{\pi} \> \left( \frac{\tau}{2} \right)^{\alpha}. 
\end{align*}
(ii) If $\alpha =1$, then
\begin{align*}
\lim_{n \to \infty} n^2 \> p_{2n} (0) = \left( \frac{\tau}{2} \right)^{2} \> \{ J_0 (\tau)^2  + J_1 (\tau)^2 \}.
\end{align*}
(iii) If $\alpha >1$, then 
\begin{align*}
\lim_{n \to \infty} n^{2 \alpha} \> p_{2n} (0) = \left( \frac{\tau}{2} \right)^{2 \alpha}. 
\end{align*}
\end{cor}

Corollaries \ref{cor1} and \ref{cor2} show a difference of the asymptotic behavior of the return probability of the two models defined by $U_{n}^{R}$ and $U_{n}^{K}$. Indeed, localization occurs for $\alpha \ge 1$ case of the model defined by $U_{n}^{R}$, but there is no localization for any $\alpha >0$ in the model defined by $U_{n}^{K}$. Furthermore, the rates of convergence of the return probability of the two models are obtained. 

\section{Proofs of Theorems {\rmfamily \ref{thm1}} and {\rmfamily \ref{thm2}}}
First we will prove Theorem {\rmfamily \ref{thm1}}. Let $x_n = (n/\tau)^{2 \alpha}$ and $A_n = |b_n| / |a_n|$. So $|a_{n}|=1/\sqrt{x_{n}}$ and $A_n = \sqrt{x_n -1} \sim \sqrt{x_n}$. We assume that $n$ is even. We begin with 
\begin{align*} 
K_n 
=
|a_{n}|^{n-2}
\sum_{\gamma =1} ^{n/2}
\left( - A_n^2  \right)^{\gamma -1} 
{n/2 -1 \choose \gamma- 1}^2  
=
|a_{n}|^{n-2}
\sum_{\gamma =0} ^{n/2-1}
\left( i A_n  \right)^{2\gamma} 
{n/2 -1 \choose \gamma}^2.  
\end{align*}
In order to compute the asymptotic behavior of $K_n$ as $n \to \infty$, we introduce 
\begin{align*} 
\Phi_n (z) = |a_{n}|^{n-2} \left( 1 + i A_n z \right)^{n/2-1} \> \left( 1 + i A_n z^{-1} \right)^{n/2-1}.
\end{align*}
Let $[z^n] \left( f(z) \right)$ denote the coefficient of $z^n$ in the formal power series $f(z) = \sum_{n=0}^{\infty} f_n z^n$, so that $[z^n] \left( f(z) \right) = f_n.$ Then we should remark that
\begin{align}
K_n = [z^0] \left( \Phi_n (z) \right).
\label{atsui1}
\end{align}
So we will calculate the asymptotic behavior of $\Phi_n (z)$ as $n \to \infty$. Put $w =z + z^{-1}$. Then we have
\begin{align*} 
\Phi_n (z) = \left\{ |a_{n}|^{2}\left( 1 + i w A_n - A_n ^2 \right) \right\}^{n/2-1} \sim (-1)^{n/2-1} \left( 1 - i w / \sqrt{x_{n}} \right)^{n/2-1}.
\end{align*}
Noting that $(n/2-1)/\sqrt{x_{n}} \sim \tau^{\alpha} n^{1- \alpha}/2$, we obtain
\begin{align} 
\Phi_n (z) \sim (-1)^{n/2-1} \exp \left( - i \tau^{\alpha} n^{1 - \alpha} \left( z +z^{-1} \right)/2 \right) .
\label{atsui2}
\end{align}
On the other hand, the generating function of the modified Bessel function is given by
\begin{align} 
\exp \left( u \left( z +z^{-1} \right)/2 \right) = \sum_{m= - \infty}^{\infty} I_m (u) z^m,
\label{atsui3}
\end{align}
see \cite{Watson1944}. By Eqs. (\ref{atsui2}), and (\ref{atsui3}), we have
\begin{align} 
[z^0] \left( \Phi_n (z) \right) \sim (-1)^{n/2-1} I_0 (- i \tau^{\alpha} n^{1- \alpha}).
\label{atsui4}
\end{align}
From $I_0 ( -i z) = J_0 (z)$, we get 
\begin{align*} 
[z^0] \left( \Phi_n (z) \right) \sim (-1)^{n/2-1} J_0 (\tau^{\alpha} n^{1- \alpha}).
\end{align*}
Therefore Eq. (\ref{atsui1}) yields 
\begin{align} 
K_n \sim (-1)^{n/2-1} J_0 (\tau^{\alpha} n^{1- \alpha}).
\label{atsui5}
\end{align}
Next we consider   
\begin{align*} 
L_n 
= |a_{n}|^{n-2}\sum_{\gamma =1} ^{n/2}
\left( - A_n^2 \right)^{\gamma -1}
\frac{1}{\gamma}  
{n/2 -1 \choose \gamma- 1}^2  
=
- \frac{2 i |a_{n}|^{n-2}}{n A_n} \> 
\sum_{\gamma =0} ^{n/2-1}
\left( i A_n  \right)^{2\gamma +1} 
{n/2 \choose \gamma + 1}
{n/2-1 \choose \gamma}. 
\end{align*}
In order to compute the asymptotic behavior of $L_n$ as $n \to \infty$, we introduce 
\begin{align*} 
\widetilde{\Phi}_n (z) = |a_{n}|^{n-2}\left( 1 + i A_n z \right)^{n/2} \> \left( 1 + i A_n z^{-1} \right)^{n/2-1}.
\end{align*}
Then we should remark that
\begin{align} 
L_n = - \frac{2 i}{n A_n} \> \left\{ [z] \left( \widetilde{\Phi}_n (z) \right) \right\}.
\label{atsui6}
\end{align}
Noting that $\widetilde{\Phi}_n (z) = \left( 1 + i A_n z \right) \Phi_n (z)$, we see that
\begin{align} 
[z] \left( \widetilde{\Phi}_n (z) \right) = [z] \left( \Phi_n (z) \right) + i A_n [z^0] \left( \Phi_n (z) \right).
\label{atsui7}
\end{align}
By Eqs. (\ref{atsui2}), and (\ref{atsui3}), we have
\begin{align*} 
[z] \left( \Phi_n (z) \right) \sim (-1)^{n/2-1} I_1 (- i \tau^{\alpha} n^{1- \alpha}).
\end{align*}
From $I_1 ( -i z) = (-i) J_1 (z)$ (see \cite{Watson1944}), we get 
\begin{align} 
[z] \left( \Phi_n (z) \right) \sim (-1)^{n/2-1} (-i) J_1 (\tau^{\alpha} n^{1- \alpha}).
\label{atsui9}
\end{align}
By using Eqs. (\ref{atsui4}), (\ref{atsui5}), (\ref{atsui6}), (\ref{atsui7}), (\ref{atsui9}), and $A_n \sim \sqrt{x_n}$, we obtain
\begin{align*} 
L_n = - \frac{2 i}{n} \> \left\{ \frac{[z] \left( \Phi_n (z) \right)}{A_n} + i [z^0] \left( \Phi_n (z) \right) \right\} \sim \frac{2}{n} (-1)^{n/2-1} \> \left\{ J_0 (\tau^{\alpha} n^{1- \alpha}) - \frac{J_1 (\tau^{\alpha} n^{1- \alpha})}{\sqrt{x_n}} \right\}.
\end{align*} 
Here we use the following result given by Konno \cite{Konno2002, Konno2005a}.
\begin{lem} 
\label{lemK1} 
When $n$ is even, we have
\begin{align*}
P(X_n=0) 
&= |a_n|^{2(n-1)} 
\sum_{\gamma =1} ^{n/2} \sum_{\delta =1} ^{n/2}
\left(- A_n^2 \right)^{\gamma + \delta} 
{n/2-1 \choose \gamma- 1}^2 
{n/2-1 \choose \delta- 1}^2  
\\
&
\qquad \qquad \qquad \qquad \times
\left\{ 
\left( \frac{n}{2} \right)^2 \frac{1}{\gamma \delta} 
- \frac{n}{2} \left( \frac{1}{\gamma} + \frac{1}{\delta} \right)
+ \frac{1}{1-|a_n|^2} 
\right\}
\\
&= |a_n|^{2} (- A_n^2)^2 \left\{ \left( \frac{n}{2} \right)^2 L_n^2 - n L_n K_n + \left( \frac{1}{1-|a_n|^2} \right) K_n^2 \right\},
\end{align*}
where $A_n = |b_n|/|a_n|.$ 
\end{lem}
By using this lemma, noting that $n$ is even, we see that
\begin{align*} 
P(X_n = 0) 
&\sim x_n \> \left\{ \left( \frac{n}{2} \right)^2 L_n^2 - n L_n K_n + \left( \frac{1}{1-1/x_n} \right) K_n^2 \right\}
\\
&\sim x_n \> (-1)^n \> \left\{ \left( J_0 - \frac{J_1}{\sqrt{x_n}} \right)^2  - 2 \left( J_0 - \frac{J_1}{\sqrt{x_n}} \right) J_0 + \left( 1 + \frac{1}{x_n} \right) J_0^2 \right\}
\\
&\sim  J_0^2 + J_1^2, 
\end{align*}
where $J_k = J_k (\tau^{\alpha} n^{1- \alpha})$ for $k=0,1$. So 
\begin{align*} 
P(X_{2n} = 0) \sim  J_0 (\tau^{\alpha} (2n)^{1- \alpha})^2 + J_1 (\tau^{\alpha} (2n)^{1- \alpha})^2
\end{align*}
completes the proof of Theorem {\rmfamily \ref{thm1}}. 

From now on we will prove Theorem {\rmfamily \ref{thm2}}. In this case, we should note that $|a_{n}|=\sqrt{1-1/x_{n}}$ and $A_n = |b_n|/|a_n| \sim 1/\sqrt{x_n}$. Similarly, we see that 
\begin{align*} 
\Phi_n (z) 
&= 
|a_{n}|^{n-2}\left( 1 + i A_n z \right)^{n/2-1} \> \left( 1 + i A_n z^{-1} \right)^{n/2-1}
\\
&= 
\left\{|a_{n}|^{2}\left( 1 + i w A_n - A_n ^2 \right)\right\}^{n/2-1} 
\\
&\sim \left( 1 + i w /\sqrt{x_n} \right)^{n/2-1}.
\end{align*}
Thus we have
\begin{align*} 
\Phi_n (z) 
\sim \exp \left(i \tau^{\alpha} n^{1 - \alpha} \left( z +z^{-1} \right)/2 \right).
\end{align*}
Then 
\begin{align*} 
K_n \sim J_0 (\tau^{\alpha} n^{1- \alpha}), \quad L_n \sim \frac{2}{n} \> \left\{ J_0 (\tau^{\alpha} n^{1- \alpha}) + \sqrt{x_n} J_1 (\tau^{\alpha} n^{1- \alpha}) \right\}.
\end{align*}
By using Lemma {\rmfamily \ref{lemK1}}, noting that $n$ is even, we see that
\begin{align*} 
P(X_n = 0) 
&\sim \left( 1 - \frac{1}{x_n} \right) \> \frac{1}{x_n^2} \> \left\{ \left( \frac{n}{2} \right)^2 L_n^2 - n L_n K_n + x_n K_n^2 \right\}
\\
&\sim \frac{1}{x_n^2} \> \left\{ \left( J_0 +  \sqrt{x_n} J_1 \right)^2  - 2 \left( J_0 + \sqrt{x_n} J_1 \right) J_0 + x_n J_0^2 \right\}
\\
&\sim \left( \frac{\tau}{n} \right)^{2 \alpha} ( J_0^2 + J_1^2 ), 
\end{align*}
where $J_k = J_k (\tau^{\alpha} n^{1- \alpha})$ for $k=0,1$. Therefore the desired conclusion is obtained.

\section{Final-time dependent classical random walk}
In this section we study the classical counterpart of our final-time dependent QW. We first consider the evolution of the time-independent (i.e., usual) CRW on the line is given by 
\begin{align*}
&
P( \mbox{ the particle moves one unit to the left})
\\
&
\qquad 
=
\left\{
\begin{array}{rl}
p, & \qquad \mbox{if the previous step was to the left,} \\
1-q, & \qquad \mbox{if the previous step was to the right,} 
\end{array}
\right.
\end{align*}
and
\begin{align*} 
&
P( \mbox{ the particle moves one unit to the right})
\\
&
\qquad 
=
\left\{
\begin{array}{rl}
1-p, & \qquad \mbox{if the previous step was to the left,} \\
q, & \qquad \mbox{if the previous step was to the right.} 
\end{array}
\right.
\end{align*}
This model is also called the correlated random walk (see \cite{Konno2009a}, for example). When $p=q$, the probability that the particle moves one unit in the same direction as the last step is $p$, and the probability that the particle moves one unit in the opposite direction as the last step is $1-p$. In particular, if $p=q=1/2$, then the walk is equivalent to the well-known symmetric (non-correlated) random walk, i.e., the particle moves at each step either one unit to the left with probability $1/2$ or one unit to the right with probability $1/2.$ The directions of different steps are independent of each other.

Here we present the definition of the final-time dependent CRW. For the general final-time dependent setting, the time evolution of the CRW is determined by a sequence of $2 \times 2$ transition matrices, $\{ V_n : n \in \ZM_{+} \}$, where
\begin{align*}
V_n =
\left[
\begin{array}{cc}
a_n & b_n \\
c_n & d_n
\end{array}
\right],
\end{align*}
with $a_n, b_n, c_n, d_n \in [0,1]$ and $a_n + c_n = b_n + d_n = 1.$ The subscript $n$ indicates the final time. In particular, we investigate the following model corresponding to the QW determined by $U^{R}_n$ with $\beta, \theta > 0$: 
\begin{align*}
V^{R}_n =
\left[
\begin{array}{cc}
\left( \frac{\theta}{n} \right)^{\beta}  & 1 - \left( \frac{\theta}{n} \right)^{\beta}  \\
1 - \left( \frac{\theta}{n} \right)^{\beta} & \left( \frac{\theta}{n} \right)^{\beta}
\end{array}
\right].
\end{align*}
Moreover we consider the following model corresponding to the QW determined by $U^{K}_n$ with $\beta, \theta > 0$: 
\begin{align*}
V^{K}_n =
\left[
\begin{array}{cc}
1 - \left( \frac{\theta}{n} \right)^{\beta}  &  \left( \frac{\theta}{n} \right)^{\beta} \\
\left( \frac{\theta}{n} \right)^{\beta} & 1 - \left( \frac{\theta}{n} \right)^{\beta}
\end{array}
\right].
\end{align*}
In this paper, we choose $\varphi_{\ast} = {}^T [1/2,1/2]$ as the initial distribution of the final-time dependent CRWs. We first give a result on the return probability of the CRW defined by $V_{n}^{R}$. 

\begin{thm}
\label{thm3}
For final-time dependent 1D CRW determined by $V_n ^{R}$ with $\beta, \theta >0$, we have 
\begin{align*}
p_{2n} (0) \sim \exp \left( - \theta^{\beta} (2n)^{1- \beta} \right) \> \left\{ I_0 (\theta^{\beta} (2n)^{1 - \beta})  + I_1 (\theta^{\beta} (2n)^{1 - \beta}) \right\},
\end{align*}
where $p_{2n} (0)$ is the return probability at the origin at the final time $2n$ and $I_k(x)$ is the modified Bessel function of the first kind of order $k$.
\end{thm}
The proof will appear in the next section. By this theorem, we have
\begin{cor}
\label{cor3}
(i) If $0 < \beta < 1$, then 
\begin{align*}
\lim_{n \to \infty} n^{(1- \beta)/2} \> p_{2n} (0) = \frac{1}{\sqrt{\pi}} \> \left( \frac{2}{\theta} \right)^{\beta/2}. 
\end{align*}
(ii) If $\beta =1$, then 
\begin{align*}
\lim_{n \to \infty} p_{2n} (0) = e^{- \theta} \> \left\{ I_0 (\theta) + I_1 (\theta) \right\}.
\end{align*}
(iii) If $\beta >1$, then 
\begin{align*}
\lim_{n \to \infty} n^{\beta -1} \> \{ 1 - p_{2n} (0) \} = \left( \frac{\theta}{2} \right)^{\beta}. 
\end{align*}
\end{cor}
From now on a proof of the corollary will be given. Concerning part (i), it is sufficient to note that as $x \to \infty$, 
\begin{align*}
I_0 (x) \sim \frac{e^x}{\sqrt{2 \pi x}} \> \left( 1 + \frac{1}{8x} \right), \quad I_1 (x) \sim \frac{e^x}{\sqrt{2 \pi x}} \> \left( 1 - \frac{3}{8x} \right),
\end{align*}
see \cite{Watson1944}. Part (ii) is trivial. As for part (iii), we observe that for small $x$, 
\begin{align*}
I_0 (x) \sim 1 + \frac{x^2}{4}, \quad I_1 (x) \sim \frac{x}{2} + \frac{x^3}{16},
\end{align*}
see \cite{Watson1944}. So we get $I_0 (x) + I_1 (x) \sim 1 + x/2$ and then $e^{-x}\left\{I_0 (x) + I_1 (x)\right\} \sim 1 - x/2$. Thus we have the desired conclusion.

Next we present a result on the return probability of the CRW given by $V_{n}^{K}$. 

\begin{thm}
\label{thm4}
For final-time dependent 1D CRW determined by $V_n ^{K}$ with $\beta, \tau >0$, we have 
\begin{align*}
p_{2n} (0) \sim \left( \frac{\theta}{2n} \right)^{\beta} \> \exp \left( - \theta^{\beta} (2n)^{1 - \beta} \right) \> \left\{ I_0 (\theta^{\beta} (2n)^{1 - \beta})  + I_1 (\theta^{\beta} (2n)^{1 - \beta}) \right\}.
\end{align*}
\end{thm}
The proof can be seen in the next section. 
From this theorem, we obtain the following result as in Corollary {\rmfamily \ref{cor3}}. So we will omit the proof.

\begin{cor}
\label{cor4}
(i) If $0 < \beta < 1$, then
\begin{align*}
\lim_{n \to \infty} n^{(\beta +1)/2} \> p_{2n} (0) = \frac{1}{\sqrt{\pi}} \> \left( \frac{\theta}{2} \right)^{\beta/2}. 
\end{align*}
(ii) If $\beta =1$, then
\begin{align*}
\lim_{n \to \infty} n \> p_{2n} (0) =  \frac{ \theta e^{- \theta}}{2} \> \{ I_0 (\theta)  + I_1 (\theta) \}.
\end{align*}
(iii) If $\beta >1$, then 
\begin{align*}
\lim_{n \to \infty} n^{\beta} \> p_{2n} (0) = \left( \frac{\theta}{2} \right)^{ \beta}. 
\end{align*}
\end{cor}

\section{Proofs of Theorems {\rmfamily \ref{thm3}} and {\rmfamily \ref{thm4}}}
First we will give a proof of Theorem {\rmfamily \ref{thm3}}. Let $y_n = (n/\theta)^{\beta}$ and $B_n = b_n / a_n$. Note that $a_{n}=1/y_{n}$ and $B_n = y_n - 1$. We assume that $n$ is even. We begin with 
\begin{align*} 
K_n^{(c)} 
=
a_{n}^{n-2}
\sum_{\gamma =1} ^{n/2}
B_n^{2(\gamma -1)} 
{n/2 -1 \choose \gamma- 1}^2  
=
a_{n}^{n-2}
\sum_{\gamma =0} ^{n/2-1}
B_n^{2\gamma} 
{n/2 -1 \choose \gamma}^2.  
\end{align*}
In order to compute the asymptotic behavior of $K_n^{(c)}$ as $n \to \infty$, we introduce 
\begin{align*} 
\Phi_n (z) = a_{n}^{n-2}\left( 1 + B_n z \right)^{n/2-1} \> \left( 1 + B_n z^{-1} \right)^{n/2-1}.
\end{align*}
Note that
\begin{align*}
K_n ^{(c)} = [z^0] \left( \Phi_n (z) \right).
\end{align*}
So we will calculate the asymptotic behavior of $\Phi_n (z)$ as $n \to \infty$. Then we have
\begin{align*} 
\Phi_n (z) = \left\{a_{n}^{2}\left( 1 + w B_n + B_n ^2 \right)\right\}^{n/2-1} \sim \left\{ 1 + (w-2)/y_{n} \right\}^{n/2-1}.
\end{align*}
Noting that $(n/2-1)/y_{n} \sim \theta^{\beta} n^{1- \beta}/2$, we obtain
\begin{align} 
\Phi_n (z) \sim \exp \left( -\theta^{\beta} n^{1 - \beta} \right) \exp \left( \theta^{\beta} n^{1 - \beta} \left( z +z^{-1} \right)/2 \right) = \exp \left( -\theta^{\beta} n^{1 - \beta} \right) \> \sum_{m= \infty}^{\infty} I_m (\theta^{\beta} n^{1- \beta}) z^m.
\label{hayato1}
\end{align}
Then we have
\begin{align} 
[z^0] \left( \Phi_n (z) \right) \sim \exp \left( -\theta^{\beta} n^{1 - \beta} \right) I_0 (\theta^{\beta} n^{1- \beta}).
\label{hayato2}
\end{align}
Therefore
\begin{align*} 
K_n^{(c)} \sim \exp \left( -\theta^{\beta} n^{1 - \beta} \right) I_0 (\theta^{\beta} n^{1- \beta}).
\end{align*}
Next we consider   
\begin{align*} 
L_n^{(c)} 
= a_{n}^{n-2}\sum_{\gamma =1} ^{n/2}
B_n ^{2(\gamma -1)}
\frac{1}{\gamma}  
{n/2 -1 \choose \gamma- 1}^2  
=
\frac{2a_{n}^{n-2}}{n B_n} \> 
\sum_{\gamma =0} ^{n/2-1}
B_n^{2\gamma +1} 
{n/2 \choose \gamma + 1}
{n/2-1 \choose \gamma}. 
\end{align*}
In order to compute the asymptotic behavior of $L_n^{(c)}$ as $n \to \infty$, we introduce 
\begin{align*} 
\widetilde{\Phi}_n (z) = a_{n}^{n-2} \left( 1 + B_n z \right)^{n/2} \> \left( 1 + B_n z^{-1} \right)^{n/2-1}.
\end{align*}
Then we should remark that
\begin{align} 
L_n^{(c)} = \frac{2}{n B_n} \> \left\{ [z] \left( \widetilde{\Phi}_n (z) \right) \right\}.
\label{hayato3}
\end{align}
Noting that $\widetilde{\Phi}_n (z) = \left( 1 + B_n z \right) \Phi_n (z)$, we see that
\begin{align} 
[z] \left( \widetilde{\Phi}_n (z) \right) = [z] \left( \Phi_n (z) \right) + B_n [z^0] \left( \Phi_n (z) \right).
\label{hayato4}
\end{align}
On the other hand, from Eq. (\ref{hayato1}) we have
\begin{align} 
[z] \left( \Phi_n (z) \right) \sim \exp \left( -\theta^{\beta} n^{1 - \beta} \right) I_1 (\theta^{\beta} n^{1- \beta}).
\label{hayato5}
\end{align}
Therefore, by Eqs. (\ref{hayato2}), (\ref{hayato3}), (\ref{hayato4}), and (\ref{hayato5}), we obtain
\begin{align*} 
L_n^{(c)} 
&\sim \frac{2}{n} \> \exp \left( -\theta^{\beta} n^{1 - \beta} \right) \> \left\{ I_0 (\theta^{\beta} n^{1- \beta}) + \frac{I_1 (\theta^{\beta} n^{1- \beta})}{B_n} \right\}
\\
&\sim \frac{2}{n} \> \exp \left( -\theta^{\beta} n^{1 - \beta} \right) \> \left\{ I_0 (\theta^{\beta} n^{1- \beta}) + \frac{I_1 (\theta^{\beta} n^{1- \beta})}{y_n} \right\}.
\end{align*} 
Here we use the following fact given by \cite{Konno2009a}.
\begin{lem} 
\label{lemK2} 
When $n$ is even, we have
\begin{align*}
P(Y_n=0) 
= a_n^{n/2-2} d_n^{n/2-2} \> 
\left\{ \frac{n}{4} (a_n + d_n) b_n c_n L_n^{(c)} + \frac{1}{2} (a_n c_n + b_n d_n) (a_n d_n - b_n c_n) K_n^{(c)} \right\}.
\end{align*}
\end{lem}
By using this lemma, noting that $n$ is even, we see that
\begin{align*} 
P(Y_n = 0) 
&\sim y_n \> \left\{ \left( 1 - \frac{2}{y_n} + \frac{1}{y_n^2} \right) \frac{n}{2} L_n^{(c)} + \left( -1 + \frac{3}{y_n} - \frac{2}{y_n^2} \right) K_n^{(c)} \right\}
\\
&\sim \exp \left( - \theta^{\beta} n^{1- \beta} \right) \left\{ I_0 (\theta^{\beta} n^{1- \beta}) + I_1 (\theta^{\beta} n^{1- \beta}) \right\}.
\end{align*}
Hence, the proof of Theorem {\rmfamily \ref{thm3}} is complete. 

Now we will move to prove Theorem {\rmfamily \ref{thm4}}. In this case, we should note that $a_{n}=1-1/y_{n}$ and $B_n = b_n/a_n =1 / (y_n - 1) \sim 1/y_n$. Similarly, we see that 
\begin{align*} 
\Phi_n (z) 
&= 
a_{n}^{n-2}\left( 1 + B_n z \right)^{n/2-1} \> \left( 1 + B_n z^{-1} \right)^{n/2-1}
\\
&= 
\left\{a_{n}^{2}\left( 1 + w B_n + B_n ^2 \right)\right\}^{n/2-1} 
\\
&\sim \left( 1 + (w-2) /y_{n} \right)^{n/2-1}.
\end{align*}
Thus we have
\begin{align*} 
\Phi_n (z) 
\sim \exp \left(-\theta^{\beta} n^{1 - \beta} \right) \exp \left(\theta^{\beta} n^{1 - \beta} \left( z +z^{-1} \right)/2 \right).
\end{align*}
Then 
\begin{align*} 
K_n^{(c)} \sim \exp \left(-\theta^{\beta} n^{1 - \beta} \right) I_0 (\theta^{\beta} n^{1- \beta}), \quad L_n^{(c)} \sim \exp \left(-\theta^{\beta} n^{1 - \beta} \right) \frac{2}{n} \> \left\{ I_0 (\theta^{\beta} n^{1- \beta}) + y_n I_1 (\theta^{\beta} n^{1- \beta}) \right\}.
\end{align*}
By using Lemma {\rmfamily \ref{lemK2}} again, we see that
\begin{align*} 
P(Y_n = 0) 
&\sim \left( 1 - \frac{1}{y_n} \right) \> \frac{1}{y_n^2} \> \left\{ \frac{n}{2} L_n^{(c)} + (y_n -2) K_n^{(c)} \right\}
\\
&\sim \exp \left(-\theta^{\beta} n^{1 - \beta} \right) \> \frac{1}{y_n^2} \> \left\{  I_0 +  y_n I_1 + (y_n -2) I_0 \right\}
\\
&\sim \exp \left( - \theta^{\beta} n^{1 - \beta} \right) \left( \frac{\theta}{n} \right)^{\beta} ( I_0 + I_1 ), 
\end{align*}
where $I_k = I_k (\theta^{\beta} n^{1 - \beta})$ for $k=0,1$. Therefore we have the desired conclusion.

\section{Discussion}
In the present paper, we have the asymptotic behavior of the return probability of the final-time dependent QWs on the line defined by $U_{n}^{R}$ and $U_{n}^{K}$ and the corresponding final-time dependent CRWs on the line defined by $V_{n}^{R}$ and $V_{n}^{K}$. Under the condition $\alpha =\beta $, the rate of convergence of return probability of the QW is square of that of the CRW. The QWs (resp. CRWs) defined by $U_{n}^{R}$ with $\alpha \geq 1$ (resp. $V_{n}^{R}$ with $\beta \geq 1$) exhibit localization. On the other hand, the QWs (resp. CRWs) defined by $U_{n}^{K}$ with any $\alpha $ (resp. $V_{n}^{K}$ with any $\beta $) do not exhibit localization. 
For the CRW defined by $V_n^{K}$ with $\beta = 1$, the corresponding weak limit theorem was obtained by \cite{Konno2009a}. Furthermore, the asymptotics of the QWs (resp. CRWs) given by $U_{n}^{R}$ with $0<\alpha \leq 1$ (resp. $0<\beta \leq 1$) correspond to weak limit theorems, (i.e. Corollary 3 (resp. (A.7))) appeared in Chisaki et al.\ \cite{ChisakiEtAl2010}.


\par
\
\par\noindent
{\bf Acknowledgment.} The second author thanks F. Alberto Gr\"unbaum for useful comments on 
an early version of this paper. This work was partially supported by the Grant-in-Aid for 
Scientific Research (C) of Japan Society for the Promotion of Science (Grant No. 21540118).

\par
\
\par

\begin{small}
\bibliographystyle{jplain}

\end{small}

\end{document}